\documentclass[journal]{IEEEtran}

\usepackage{multirow}
\usepackage{amsmath}

\DeclareMathOperator*{\argmin}{Argmin}
\usepackage{graphicx}
\usepackage{amssymb}
\usepackage{booktabs}
\usepackage{acronym} 

\newacro{ml}[ML]{machine learning}
\newacro{dl}[DL]{deep learning}
\newacro{vlm}[VLM]{Vision-Language Model}
\newacro{llm}[LLM]{Large Language Model}
\newacro{mha}[MHA]{MultiHead Attention}
\newacro{dnn}[DNN]{Deep Neural Network}
\newacro{lmm}[LMM]{Large Multimodal Model}
\newacro{roi}[ROI]{Region Of Interest}
\newacro{vit}[ViT]{VIsion Transformer}
\newacro{sbert}[SBERT]{Sentence BERT}
\newacro{use}[USE]{Universal Sentence Encoder}
\newacro{prm}[PRM]{Patch Representation Misalignment}
\newacro{mpnet}[MPNET]{ Masked and Permuted Pre-training for Language Understanding}
\newacro{ssim}[SSIM]{Structural SIMilarity}
\newacro{lpips}[LPIPS]{Learned Perceptual Image Patch Similarity}
\newacro{ai}[AI]{artificial intelligence}
\newacro{nlp}[NLP]{Natural Language Processing}
\newacro{vqa}[VQA]{Visual Question Answering}
\newacro{qava}[QAVA]{Query-Agnostic Visual Attack}

\usepackage{algorithm}
\usepackage{algpseudocode}
\usepackage{soul}

\usepackage{comment}

\usepackage[most]{tcolorbox} 

\tcbset{
  colback=gray!10,    
  colframe=black,     
  boxrule=0.5pt,      
  arc=4pt,            
  left=6pt, right=6pt, top=6pt, bottom=6pt, 
}
\usepackage{pifont}
\usepackage{bm}

\usepackage{cite}

\usepackage{hyperref}
\hypersetup{
    colorlinks=true,
    linkcolor=blue,
    filecolor=magenta,      
    urlcolor=cyan,
}

\begin{document}

\title{VIP: Visual Information Protection through Adversarial Attacks on Vision-Language Models}

\author{Hanene F. Z. Brachemi Meftah, Wassim Hamidouche, Sid Ahmed Fezza, and Olivier Déforges
\thanks{This work is fully funded by Région Bretagne (Brittany region), France,
CREACH Labs and Direction Générale de l’Armement (DGA).\\}
\thanks{Hanene Brachemi Meftah and Olivier Déforges  are affiliated to Univ. Rennes, INSA Rennes, CNRS, IETR - UMR 6164, Rennes, France (e-mail: \href{mailto:hanene.brachemi@insa-rennes.fr}{hanene.brachemi@insa-rennes.fr} and \href{mailto:olivier.deforges@insa-rennes.fr}{olivier.deforges@insa-rennes.fr} ).}
\thanks{Wassim Hamidouche is affiliated to Technology Innovation Institute P.O.Box: 9639, Masdar City Abu Dhabi, UAE (e-mail: \href{mailto:whamidouche@gmail.com}{whamidouche@gmail.com}).}
\thanks{Sid Ahmed Fezza is affiliated to National Higher School of Telecommunications and ICT, Oran, Algeria (e-mail: \href{mailto:sfezza@ensttic.dz}{sfezza@ensttic.dz}).}
}

\markboth{}{}

\maketitle

\begin{abstract}
Recent years have witnessed remarkable progress in developing \acp{vlm} capable of processing both textual and visual inputs. These models have demonstrated impressive performance, leading to their widespread adoption in various applications.
However, this widespread raises serious concerns regarding user privacy, particularly when models inadvertently process or expose private visual information.
In this work, we frame the preservation of privacy in \acp{vlm} as an adversarial attack problem. We propose a novel attack strategy that selectively conceals information within designated \acp{roi} in an image, effectively preventing \acp{vlm} from accessing sensitive content while preserving the semantic integrity of the remaining image.
Unlike conventional adversarial attacks that often disrupt the entire image, our method maintains high coherence in unmasked areas. Experimental results across three state-of-the-art \acp{vlm} namely LLaVA, Instruct-BLIP, and BLIP2-T5 demonstrate up to 98\% reduction in detecting targeted \acp{roi}, while maintaining global image semantics intact, as confirmed by high similarity scores between clean and adversarial outputs.
We believe that this work contributes to a more privacy-conscious use of multimodal models and offers a practical tool for further research, with the source code publicly available at \href{https://github.com/hbrachemi/Vlm_defense-attack}{https://github.com/hbrachemi/Vlm\_defense-attack}.
\end{abstract}
 
\acresetall
\section{Introduction}
\label{sec:intro}
\Acp{vlm} have emerged as a powerful paradigm in artificial intelligence, seamlessly integrating visual and textual information to achieve remarkable performance in various tasks such as image captioning, visual question answering, and document understanding ~\cite{achiam2023gpt,team2023gemini,agrawal2024pixtral,li2024aria,wang2024qwen2}. This led to their rapid adoption in numerous applications, including content creation, customer service, and information retrieval. 
However, the widespread use of \acp{vlm} raises critical concerns regarding the privacy and security of user data. The ability of these models to accurately interpret and extract information from images poses potential risks, particularly when images contain sensitive or personal details. For instance, \acp{vlm} employed in social media platforms or surveillance systems could inadvertently expose private information or enable unauthorized tracking of individuals. These risks become especially severe when \acp{vlm} continuously learn from user-provided online data, increasing not only the risk that the model retains private information but also makes it vulnerable to privacy attacks aimed at extracting sensitive data~\cite{wang2022reconstructing,zhang2024effective,carlini2021extracting}.
Despite the increasing reliance on  \acp{vlm}, these privacy concerns remain largely unaddressed by the research community. 

Recent research has revealed that \acp{vlm} are vulnerable to adversarial attacks~\cite{carlini2024aligned,shayegani2024jailbreak,cui2024robustness}. These attacks typically fall into two categories: ``jailbreak" attacks that aim to bypass safety measures to generate `harmful' content, and accuracy-compromising attacks that disrupt the model's ability to interpret inputs correctly. Various studies have demonstrated these vulnerabilities, including attacks that manipulate captions~\cite{luo2024image}, disrupt feature extraction~\cite{hu2024firm}, and even increase energy consumption~\cite{gao2024inducing}. On the other hand, earlier works~\cite{10.1007/978-3-031-19781-9_4,advforsocialgood} have explored leveraging adversarial attacks for `social good', such as protecting users' photos from GAN-based manipulation or preserving the privacy of social media attributes. 
However, the challenge associated with these existing adversarial attacks is that they completely alter the semantic content of the input, creating a trade-off between maintaining the input's utility and ensuring privacy.

In this paper, we raise the following research question: 
\begin{tcolorbox}
``\textit{Can sensitive information in an image be selectively protected without degrading the overall image  quality or its extracted semantic content?}".
\end{tcolorbox}

We address privacy preservation in \acp{vlm} by framing it as an adversarial attack problem and propose a novel approach that conceals sensitive information within images from \acp{vlm}. 
Our method selectively masks \acp{roi} within an image, preventing the model from accessing and interpreting the sensitive data contained in those regions. This ensures privacy while preserving the overall integrity and usability of the image. 
Specifically, our approach focuses on manipulating both the attention mechanism and values matrices within the \acp{vlm}'s visual encoder by carefully crafting an adversarial perturbation that guides the model's attention away from the \ac{roi}. This effectively blinds the \ac{vlm} to the sensitive information contained within that region. The targeted manipulation of attention weights and value matrices allows us to achieve information concealment without significantly altering the overall visual content of the image.

Experimental results demonstrate the effectiveness of this approach in protecting user privacy on three different \ac{vlm} models. We observe a significant reduction of up to 98\% in the ability of \acp{vlm} to detect objects within the targeted regions. Furthermore, the descriptions generated for clean and adversarial images exhibit comparable similarity scores, indicating that our attack preserves the overall consistency of the visual content outside the masked regions.

\section{Related works}
\label{sec:relwork}
\subsection{\Acl{vlm}s}
The field of multimodal learning is witnessing a surge in models designed to integrate diverse modalities~\cite{achiam2023gpt,team2023gemini,agrawal2024pixtral,li2024aria,wang2024qwen2}, with images and text being the most prevalent. This integration enables models to tackle a wider range of tasks, including answering questions about images, generating descriptions, and even solving problems presented visually.  \Acp{vlm} exemplify this trend, combining visual and textual information to achieve these capabilities. A typical \ac{vlm} architecture comprises three main components: a pre-trained visual encoder, a projector model to align the visual and textual representations, and a pre-trained \ac{llm} decoder. While the specific decoder and fusion techniques may vary, variants of CLIP's visual encoder are widely adopted due to their effectiveness in extracting meaningful visual features. Several popular open-source \acp{vlm} follow this architectural pattern, including LLaVA~\cite{li2023m}, Instruct-BLIP~\cite{dai2023instructblip}, BLIP-2~\cite{li2023blip}, and mini GPT-4~\cite{zhu2023minigpt}.  LLaVA utilizes CLIP's~\cite{radford2019language} visual encoder to generate image features, which are then projected as input to the Vicuna \ac{llm}~\cite{chiang2023vicuna}. Instruct-BLIP, BLIP-2, and mini GPT-4 employ the EVA-CLIP~\cite{fang2023eva} visual encoder with a Q-Former for multimodal fusion while offering flexibility in utilizing different \acp{llm}, such as Vicuna and Flan-T5~\cite{chung2024scaling}. Although newer \acp{vlm} (e.g. DeepSeek~\cite{liu2024deepseek}, Llama 3 Herd~\cite{dubey2024llama}, Molmo and PixMo~\cite{deitke2024molmo}, NVIDIA's NVLM~\cite{dai2024nvlm}, 
Qwen2-VL~\cite{wang2024qwen2}, Pixtral 12B~\cite{agrawal2024pixtral},
MM1.5~\cite{zhang2024mm1}, and Emu3~\cite{wang2024emu3}
) offer enhanced performance and diverse architectures, the aforementioned models continue to serve as foundation models and remain prevalent in evaluating adversarial attacks within the current literature.  This prevalence is likely due to their established benchmarks and widespread adoption in the research community. 

\subsection{Vulnerability of VLMs to adversarial attacks}
The concept of adversarial attacks within the context of image classification was initially proposed by Szegedy~\textit{et al.}~\cite{szegedy2013intriguing}, who demonstrated the feasibility of deceiving a \ac{dl} classifier by introducing carefully crafted perturbations into the input image.  Nguyen~\textit{et al.}~\cite{nguyen2015deep} further pioneered the formalization of adversarial attacks as an optimization problem. These seminal works on adversarial examples served as a foundational milestone, revealing inherent vulnerabilities within \acp{dnn}. Consequently, numerous adversarial attacks have been developed, targeting \acp{dnn} in general~\cite{goodfellow2014explaining,byrd1995limited,c&w,chen2017zoo} and, more recently, \acp{vlm}, including targeted, untargeted, energy-latency, and jailbreak attacks.

Attacks targeting model accuracy aim to disrupt the \ac{vlm}'s ability to correctly interpret and respond to inputs. For instance, Cui~\textit{et al.}~\cite{cui2024robustness} conducted a comprehensive study investigating the robustness of \acp{vlm} against established computer vision attacks~\cite{c&w,madry2018towards,apgd}. Luo~\textit{et al.}~\cite{luo2024image} introduced the CroPA framework, a targeted attack that leverages learnable prompts to generate adversarial images capable of misleading the model on different textual inputs. Zhao~\textit{et al.}~\cite{zhao2024evaluating} employed a text-to-image generator to craft adversarial examples that align with specific target captions. Hu~\textit{et al.}~\cite{hu2024firm} developed the \ac{prm} attack, a non-targeted approach that disrupts the hierarchical features extracted by the vision encoder. By minimizing the similarity between features from clean and adversarial images, this attack leads to incorrect model outputs across various downstream tasks. 
More recently, Silva~\textit{et al.}~\cite{silva2025attacking} proposed a non-targeted adversarial attack that leverages the attention matrices to distract the \ac{vlm}'s vision encoder from the regions it attends to in the clean image. Their method also optimizes the embedding vectors produced by the visual encoder to diverge from those of the original inputs.
Zhang~\textit{et al.}~\cite{zhang2025qava} introduced the \ac{qava} attack, a non-targeted method that exploits the output of the Q-former module. Their approach samples a set of random questions and optimizes the adversarial image such that the Q-former’s output diverges significantly from that produced on the clean image. This enables the generated adversarial image to generalize over a larger set of queries.

However, these attacks either force the model to predict a fixed target output or produce entirely incorrect results by disrupting the image's semantic information, ultimately compromising its utility. In contrast, our objective is to leverage adversarial attacks for social good by selectively concealing specific information within an image while preserving its overall utility and the model's functionality.

\begin{figure}[h!]
    \centering
    \includegraphics[width=1\linewidth]{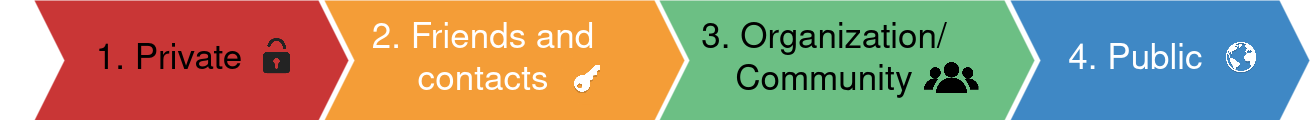}
    \caption{Privacy as a spectrum: images fall into different exposure categories.}
    \label{fig:privacy-chart}
\end{figure}

In our study, we regard user privacy as a spectrum rather than a binary attribute where different levels of data exposure are acceptable, depending on the context. 
Some images are not strictly confidential, yet their visibility should be restricted to specific social circles similarly to the audience controls on social media platforms. 
To illustrate this, we categorize privacy exposure into four levels presented in Fig.~\ref{fig:privacy-chart}.

However, this nuanced view of privacy faces a critical challenge with the growing use of \acp{vlm}: information from images belonging to the first three categories risks being accidentally exposed to individuals from the public category, compromising their intended level of privacy.
While basic methods such as blurring or masking could be viable for fully private images, they are impractical for content that still needs to be interpretable within a specific group.
On the other hand, non-private information present in the image can still be valuable for various applications, including image captioning and \ac{vqa}. 

This motivates the use of an adversarial attack that selectively conceals sensitive information within a designated \ac{roi} in an image while preserving the integrity of the image outside the \ac{roi} and maintaining its overall perceptual quality.
\section{Proposed method}
\label{sec:proposed}
\subsection{Problem statement}

\begin{figure}[t!]
    \includegraphics[width=\linewidth]{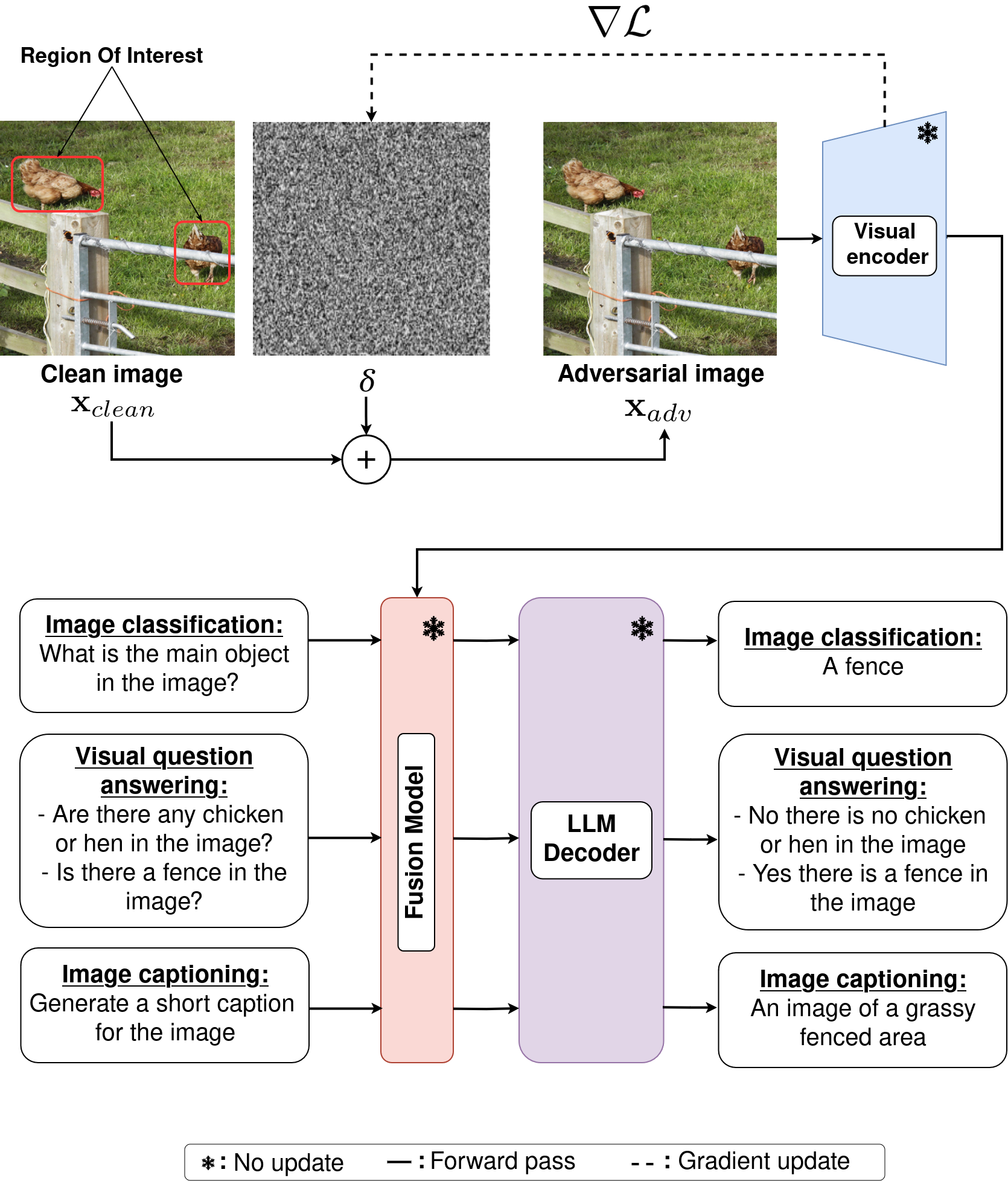}
    \caption{Overview of our proposed attack. We separate the different \ac{vlm}'s components to better illustrate our attack. Our attack uses only the visual encoder to compute the new attention maps and update the perturbation $\delta$ using loss gradients (loss is given by equations~\eqref{eqn:optim_prob1} and~\eqref{eqn:optim_prob2}). Bottom part shows expected \ac{vlm} generated outputs when inferring our resulting adversarial image $\mathbf{x}_{adv}$ with different text inputs}
    \label{fig:proposed}
\end{figure}

To achieve our objective, we leverage the model's visual encoder to iteratively optimize an image perturbation, denoted as $\delta$.
This perturbation is designed to minimize the attention assigned to image patches within the \ac{roi} in early attention blocks of the encoder. 
By reducing the attention given to these patches, we effectively prevent the information within the \ac{roi} from being propagated to subsequent layers, ultimately concealing it from the final visual features extracted by the model. An overview of the proposed attack framework is presented in Fig.~\ref{fig:proposed}.

\subsection{Threat model}
This section presents a high-level formulation of our attack's optimization problem, providing an overview of our objective and laying the groundwork for a more detailed explanation in Section~\ref{sec:prelim_design}.

Generating an adversarial example for a \ac{vlm}, denoted as $\mathcal{M}$, typically involves optimizing an adversarial perturbation $\delta$ under a constraint on its magnitude,  $\lVert \delta \rVert_p <\epsilon$. 
Let $\mathcal{T} = \{t_1,t_2,...,t_K\}$ represent a set of textual prompts and $\mathcal{R} = \{r_1,r_2,..,r_N\}$ denote a set of textual descriptors characterizing the \ac{roi} of the clean image $x_{clean}$. The goal of our attack is to generate an adversarial perturbation, $\delta$ , such that for every prompt  $t \in \mathcal{T}$, the model produces accurate outputs for the adversarial input image ($\mathbf{x}_{adv} =\mathbf{x}+\delta$) while simultaneously discarding any information $r \in \mathcal{R}$ related to the \ac{roi}.

This attack objective can be formulated as a min-max optimization problem expressed in Equation~\eqref{eqn:minmax}.  Our objective function aims to maximize the global similarity between the model's outputs for the clean and adversarial images while minimizing the similarity between the adversarial output and any textual descriptor of the \ac{roi}.
\begin{equation}
\label{eqn:minmax}
\begin{split}
    &\forall t \in \mathcal{T}:\\
    &\forall r \in \mathcal{R}:\\
    &\min_\delta  \mathcal{L}\Bigl(\mathcal{M}\bigl(\textbf{x}_{clean},t\bigl),\mathcal{M}\bigl((\textbf{x}_{clean}+\delta),t\bigl)\Bigl)\\  
    & - \lambda\;
    \mathcal{L}\Bigl(\mathcal{M}\bigl((\textbf{x}_{clean}+\delta),t\bigl),r\Bigl).
\end{split}  
\end{equation}
Here, $\mathcal{L}$ denotes a language modeling loss function that measures the semantic distance between two sentences, and  $\lambda \in \mathbb{R}^{+}$ is a Lagrangian multiplier used to balance the multiple optimization objectives.

\subsection{Preliminaries and design overview}
\label{sec:prelim_design}
Self-attention mechanisms lie at the heart of \acp{vit}~\cite{vaswani2017attention}, which are employed in various visual encoders, including CLIP and EVA-CLIP. This approach decomposes the input image into a sequence of patches, enabling the model to attend to different image regions in parallel and capture complex dependencies between them. This is achieved by computing a scaled dot-product attention over multiple heads, concatenating the resulting attention matrices, and applying a linear transformation to the concatenated output. Within a single attention head, denoted as  $h$, at layer $l$, the input to the \ac{mha} block is projected into three distinct spaces—query, key, and value—using three learnable weight matrices.  The resulting matrices are denoted as $Q$, $K$, and $V$, respectively. These matrices are then used to compute a scaled dot-product attention, as shown in Equation \eqref{eqn:dotatt}.
\begin{equation}
    \label{eqn:dotatt}
    \begin{split}
    \text{Attention}(Q,K,V) =\text{softmax}(\frac{QK^T}{\sqrt{d_k}}) V,  
    \end{split}
\end{equation}
where the term $\frac{1}{\sqrt{d_k}}$  is a scaling factor that prevents the dot product from growing too large~\cite{vaswani2017attention}.
  
In this study, we focus on the attention weights matrix, denoted as $ \mathcal{A} =\text{softmax}(\frac{QK^{T}}{\sqrt{d_k}})$. This matrix, $\mathcal{A}$, provides the attention weights assigned by token $i$ to token $j$. By manipulating these attention maps within the model's visual encoder, we can selectively restrict the information propagated from specific tokens to subsequent layers. This targeted approach eliminates the need to access other components of the \ac{vlm}, such as the fusion module or the \ac{llm}, and avoids altering the information conveyed by other tokens.

Based on this principle, our attack strategy can be formulated as the following optimization problem:
\begin{equation}
\label{eqn:optim_prob1}
\begin{split}
&\mathbf{x}_{adv} = \mathbf{x}_{clean} + \delta,\\
&\delta = \argmin_{\mathbf{x}} \sum_{l = 1}^{\text{L}_{max}} \sum_{h=1}^{H} \sum_{i=1}^{|\mathcal{S}_I|} \sum_{j \in \mathcal{S}_{ROI}} \mathcal{A}^{(l,h)}(i,j), 
\end{split}  
\end{equation}
where $\text{L}_{max}$ is a hyperparameter denoting the maximum number of \ac{mha} blocks considered during the optimization of $\delta$. $H$ represents the total number of attention heads within each \ac{mha} block. $\mathcal{S}_I$ denotes the set of indices corresponding to the complete sequence of image patches (including the CLS token), while $\mathcal{S}_{ROI} \subseteq \mathcal{S}_I$ represents the subset of indices corresponding to patches within the \ac{roi}.  Finally, $\mathcal{A}^{(l,h)}$ denotes the attention weighs matrix computed at the $h$-th head of the $l$-th \ac{mha} block.

\begin{figure}
    \includegraphics[width=\linewidth]{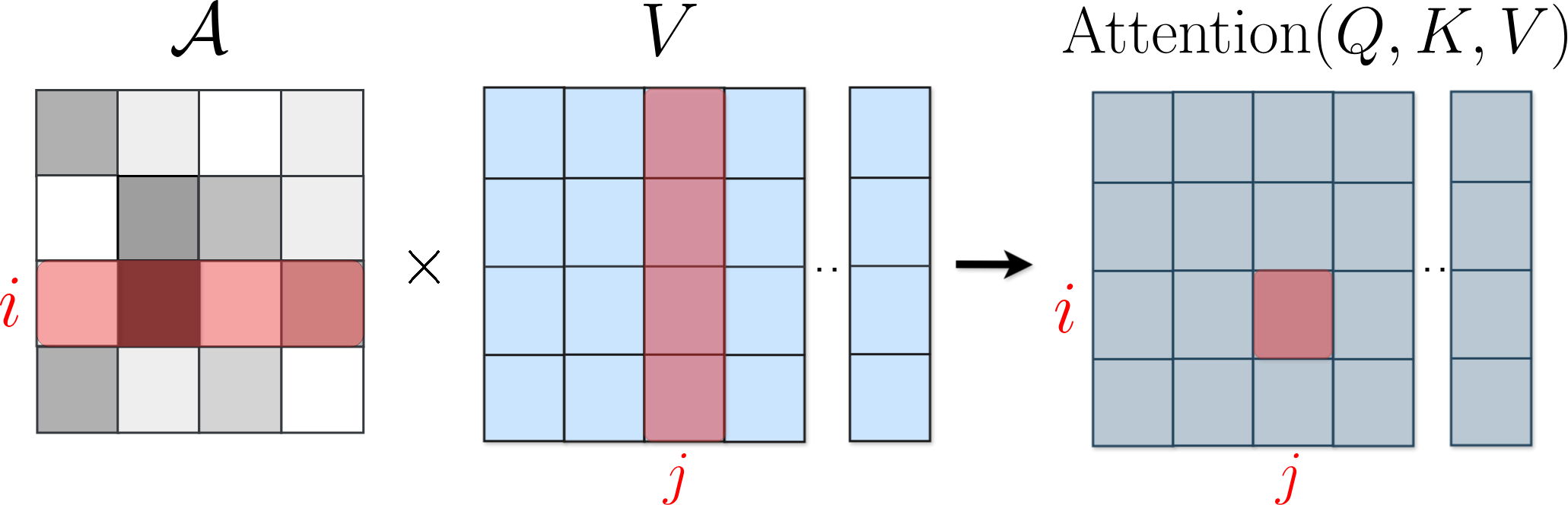}
    \caption{Illustration of the scaled-dot product attention on a single head level.}
    \label{fig:attention}
\end{figure}

Furthermore, as expressed in Equation~\eqref{eqn:dotatt}, each element $(i,j)$ of the resulting attention matrix, $\text{Attention}(Q,K,V)$, is a weighted sum over the $j$-th column of $V$, with weights determined by the $i$-th row of $\mathcal{A}$. Figure~\ref{fig:attention} illustrates this scaled dot-product attention mechanism.
Therefore, we can further attenuate the information conveyed by specific tokens by minimizing their corresponding values in the value matrix, $V$.  This leads to a refined optimization problem, incorporating an additional term that targets the value matrices, as expressed in Equation~\eqref{eqn:optim_prob2}.
\begin{equation}
\label{eqn:optim_prob2}
\begin{split}
\mathbf{x}_{adv} = &\mathbf{x}_{clean} + \delta,\\
\delta = \argmin_{\mathbf{x}} &\sum_{l = 1}^{\text{L}_{max}} \sum_{h=1}^{H} \left( \sum_{i=1}^{|\mathcal{S}_I|} \sum_{j \in \mathcal{S}_{ROI}} \mathcal{A}^{(l,h)}(i,j)\right.\\
& \left.+ \lambda_v \sum_{j \in \mathcal{S}_{ROI}} \sum_{k=1}^{d_{v}}  \rVert V^{(l,h)}(j,k)\lVert_2 \right),
\end{split}  
\end{equation}
where $\lambda_v \in \mathbb{R}^{+}$ is a Lagrangian multiplier, $d_{v}$ represents the dimension of the value matrix, and $V^{(l,h)}$
denotes the value matrix extracted from the $h$-th head of the $l$-th \ac{mha} block. The pseudocode of our method, along with a detailed explanation, is provided in subsection~\ref{supp:pseudocode}.

\subsection{Pseudocode}
\label{supp:pseudocode}
\begin{algorithm*}
\caption{Adversarial attack against \acp{vlm} (VIP)}\label{alg:algo}
\begin{algorithmic}[1]
\Require Clean image $\mathbf{x}_{clean}$, Vision-Language Mode $\mathcal{M}$, Region Of Interest $boxes$, $\text{L}_{max}$, $\lambda_v$, a learning rate $\alpha$ and a number of maximal iterations $N$
\State $\mathcal{M}_v \gets \text{ExtractVisualEncoder}(\mathcal{M})$
\State $H,W,patch\_dim \gets \text{GetInputParams}(\mathcal{M}_v)$
\State $Seq\_len \gets \lceil\frac{HW}{patch\_dim^2}\rceil+1$ \Comment{A CLS token is inserted at the start of the sequence of patches}
\State $\mathcal{S}_{ROI} \gets \text{ExtractROITokenIdx}(boxes,\mathbf{x}_{clean},patch\_dim)$

\State $\delta_0 \gets 0$
\Ensure IsNotEmpty($\mathcal{S}_{ROI}$)
\For{$t \gets 0$ to (N-1)}

\State $\mathcal{L}_{att} \gets 0$ 
\State $\mathcal{L}_{V} \gets 0$
\State $\mathcal{L}_t \gets 0$

\State $\mathbf{x}_{adv} \gets \mathbf{x}_{clean} + \delta_{t}$ 
\State $\mathbf{x}_{adv} \gets \text{Clip}(\mathbf{x}_{adv},0,255)$ \Comment{project adversarial image to valid image range}

\State $\mathcal{H} \gets  \text{ExtractHiddenStates}(\mathcal{M}_v(\mathbf{x}_{adv}))$

\For{$l \gets 1 $ to $\text{L}_{max}$}   
\For{$h \gets 1 $ to $H$}   

\State $\mathcal{A} \gets \mathcal{H}(``Attention",l,h)$
\State $V \gets \mathcal{H}(``V",l,h)$

\For{$j \in \mathcal{S}_{ROI}$}   
\For{$i \gets 1$ to $Seq\_len$}   
\State $\mathcal{L}_{att} \gets \mathcal{L}_{att} +  \mathcal{A}(i,j)$
\EndFor
\State $\mathcal{L}_{V} \gets \mathcal{L}_{V} + \lVert{V(j)}\rVert_2$
\EndFor
\EndFor
\EndFor

\State $\mathcal{L}_{t} \gets \mathcal{L}_{att} +\lambda_v \mathcal{L}_{V}$
\State $\delta_{t+1} \gets \delta_{t} - \alpha \;\text{Sign} (\nabla_{\delta_{t}} \mathcal{L}_t)$ 
\EndFor
\end{algorithmic}
\end{algorithm*}
In algorithm~\ref{alg:algo}, we present a simplified version of our proposed adversarial attack algorithm against large \acp{vlm}. We emphasize that several procedures we use are pre-implemented and available within the Hugging Face framework, although they differ slightly from how we present them in the algorithm due to their object-oriented structure. This includes functions such as \texttt{ExtractVisualEncoder} and \texttt{ExtractHiddenStates}. The \texttt{Clip} procedure (not to be confused with the CLIP model) used in line 11 is also provided by the PyTorch framework. While the procedure \texttt{ExtractROITokenIdx} iterates through all the extracted patches from the image and checks whether they are located in the \ac{roi} or not, it then returns a list of indices corresponding to the patches that are in the ROI (the index of the list starts from 1 instead of 0 in order to account for the CLS token). For this purpose, the procedure requires parameters such as the model input's dimensions, the patch's dimensions, and the box's coordinates. We also highlight that although our algorithm may seem complex at first glance, it is actually very efficient in practice. Thanks to the PyTorch framework, we minimize the use of loops, relying instead on its efficient optimization of matrix operations. To further support this claim, we report the average run times required to generate our adversarial attack using different configurations in subsection~\ref{supp:exe_time}.

\section{Experimental setup}
\subsection{Models}
To evaluate the effectiveness of our attack, we selected three prominent \acp{vlm}: LLaVA 1.5 combined with the Vicuna 7B \ac{llm}, BLIP-2 integrated with the Flan-T5 XL \ac{llm} and Instruct-BLIP also employing the Vicuna 7B \ac{llm}. These models are widely used in the evaluation of adversarial attacks~\cite{cui2024robustness,hu2024firm,zhao2024evaluating,carlini2024aligned} and exhibit distinct architectural configurations. For brevity, we will refer to these models as LLaVA, BLIP2-T5 and Instruct-BLIP respectively, throughout the remainder of this paper.

\subsection{Dataset}
We evaluate the effectiveness of our attack on a subset of 1000 images randomly selected from the ImageNet validation set~\cite{russakovsky2015imagenet}. This dataset is particularly suitable for our evaluation as it provides a diverse collection of images with corresponding bounding box annotations. We leverage these bounding boxes as \acp{roi}, with the objective of our attack being to prevent the model from detecting and describing the object within the \ac{roi}. We also ensured that all \acp{roi} within these samples are correctly detected in the absence of attack.

\subsection{Attacks}
Due to the absence of well-established baseline attacks for a fair comparison, we evaluate the effectiveness of our proposed attack strategy against the \ac{prm} attack~\cite{hu2024firm}. The \ac{prm} attack uses patch-wise embedding optimizations, which allow for more localized modifications. As a result, their approach can be adapted to our setting by targeting only patches in the \acs{roi} instead of all image patches.
We integrated a minor adaptation to the \ac{prm} attack to better align with our problem setting by restricting the patches whose features are targeted by the attack to those located within the \ac{roi}.

In our approach, we prioritize attack effectiveness and minimal visual disruption. Therefore, we do not enforce a perturbation norm constraint in our primary experiments. However, we present an ablation study with additional results using norm constraints in subsection~\ref{supp:Lp} for a comprehensive analysis. To ensure a fair comparison, we evaluate the \ac{prm}  method both with and without the standard constraint, denoting the unconstrained variant as \ac{prm}-U.
\subsection{Evaluation metrics}
The success of our attack is measured by two key factors: the failure to detect the object within the \ac{roi} and the preservation of background information. The first factor is assessed by computing the detection accuracy of the object within the \ac{roi}.  To evaluate this, we prompt the encoder-decoder \ac{vlm} with the adversarial image and the following textual input: `\textit{Is there any [label] in the image?}'. This prompting strategy is consistent with findings in~\cite{cui2024robustness}, which indicate that incorporating additional context into the model's text input enhances its resilience to adversarial attacks. Their experimental results suggest that a \ac{vlm} is more likely to produce a correct classification response when presented with a visual question (e.g. a yes/no question) rather than alternative prompts like `A photo of:' or `What is the main object in this image? Short answer:'. Consequently, our goal is to develop an adversarial attack that reliably prevents the model from detecting the object within the \ac{roi}, despite the presence of these conditions.

To assess the preservation of background information, we compute similarity scores at both the image embedding level and the generated text level. At the image level, we utilize the \ac{vlm}'s image encoder to extract high-level features from both the clean and adversarial images. We then calculate the cosine similarity between these feature representations as a measure of similarity. Similarly, at the text level, we compute the cosine similarity of the encoded captions generated by the \ac{vlm} for both the original and adversarial images. For this purpose, we utilize three sentence encoder transformers: \ac{sbert}~\cite{reimers2019sentence}, the \ac{use}~\cite{cer2018universal} and the \ac{mpnet}~\cite{song2020mpnet}. These encoders are specifically designed to generate high-quality sentence embeddings and are widely employed for semantic textual similarity tasks.


In addition to these metrics, we report the average Structural Similarity Index Measure (SSIM) and Learned Perceptual Image Patch Similarity (LPIPS) scores computed on the adversarial examples, where a higher SSIM score and a lower LPIPS score indicate higher perceptual similarity between the clean and adversarial images.


\subsection{Implementation details}
We utilize the Adam optimizer ~\cite{kingma2014adam} with a learning rate of $\alpha = 10^{-3}$ for our attack. In Equation \eqref{eqn:optim_prob2}, we set  $\lambda_v$ to 1. The number of attackable layers, $\text{L}_{max}$ is set to 1, 24, and 24 for the LLaVA, BLIP2-T5, and Instruct-BLIP models, respectively. 
For the \ac{prm}, we utilize their reported attack parameters for both constrained and unconstrained settings. For both attacks, we set the maximum number of iterations at 1500 and employ an early stopping strategy with a patience parameter of 10.  Additionally, a convergence test is performed every 100 iterations to evaluate the attack's progress. The attack terminates automatically once the non-detection objective is satisfied.




\section{Results and analysis}
\subsection{Attack effectiveness}

To assess the effectiveness of our attack, we first focus on the detectability of objects within the \ac{roi} and the similarity between captions generated for clean and adversarial images. 
Table~\ref{tab:ASR} reports the detection accuracy and the perceptual quality scores of the adversarial examples generated on the three considered models: LLaVA, Instruct-BLIP and BLIP2-T5. 

\begin{table}[h!]
\caption{Reported detection percentages of objects within the \protect\ac{roi} computed on 1000 samples of the ImageNet validation dataset alongside the average adversarial images' quality scores. Entries in bold indicate the highest scores obtained. 
}
\label{tab:ASR}
\resizebox{\columnwidth}{!}{%
\begin{tabular}{llccc}
\hline
\multirow{2}{*}{VLM}           & \multirow{2}{*}{Attack} & Detection (\%)  ↓                                   & \multicolumn{2}{c}{Perceptual fidelity} \\ \cline{4-5} 
                       &                  &                                                              & \acs{ssim} ↑ & \acs{lpips} ↓ \\ \hline
\multirow{4}{*}{LLaVA} & \ac{prm}  &  17.10     & 0.986       & 0.006        \\
                       & \ac{prm}-U              & 13.46           & 0.670  & 0.360   \\
                       & Ours (A)         & 07.00           & 0.818  & 0.125   \\
                       & Ours (A+V)       & \textbf{01.70}           & 0.769  & 0.184   \\ \hline
\multirow{4}{*}{Instruct-BLIP} & \ac{prm}         & 34.10  & 0.987              & 0.005              \\
                       & \ac{prm}-U              & 17.50           & 0.777  & 0.237   \\
                       & Ours (A)         & 28.60           & 0.838  & 0.152   \\
                       & Ours (A+V)       & \textbf{10.50}  & 0.864  & 0.127   \\ \hline
\multirow{4}{*}{BLIP2-T5}      & \ac{prm}         & 19.60  & 0.987              & 0.005              \\
                       & \ac{prm}-U              & 14.30           & 0.776  & 0.238   \\
                       & Ours (A)         & 14.10           & 0.895  & 0.093   \\
                       & Ours (A+V)       & \textbf{06.10}  & 0.893  & 0.097   \\ \hline
\end{tabular}%
}
\end{table}

\begin{figure}
    \includegraphics[width=1\linewidth]{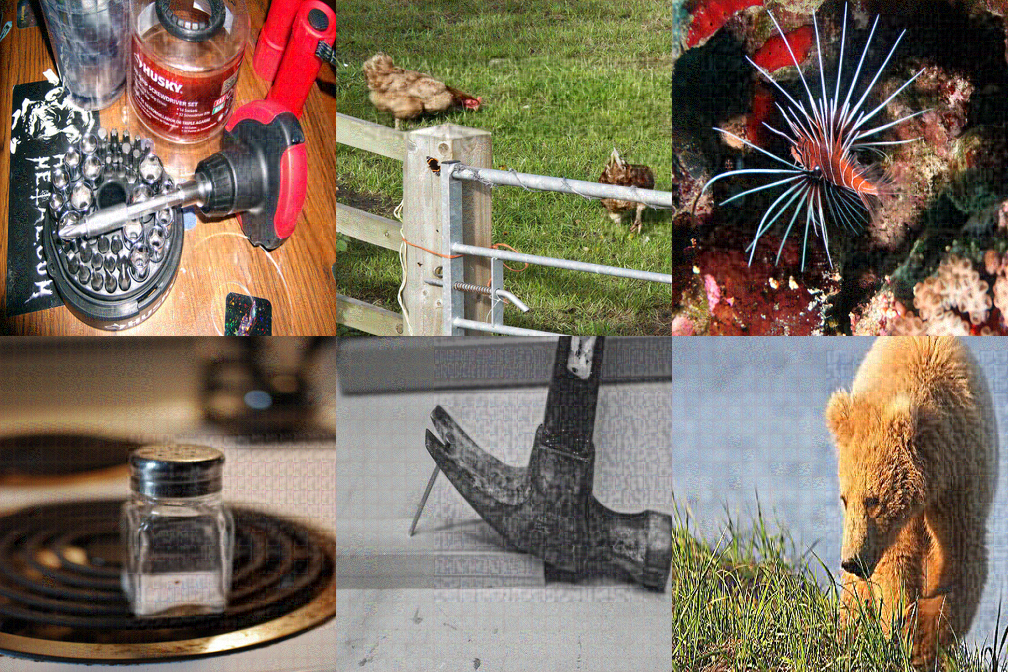}
    \caption{A perceptual illustration of the perturbation's visibility on a few adversarial samples generated against the LLaVA \ac{vlm} under the `A+V' setting.
    Adversarial perturbation tends to be more perceptible in images with flat or uniform regions, such as those presented in the second row, and is less detectable in areas with high texture, as observed in the images of the first row.}
    \label{fig:perturbation}
\end{figure}

The results presented in Table~\ref{tab:ASR} show that our attacks significantly reduce the ability of various \acp{vlm} to identify objects within the \ac{roi} decreasing significantly the detection accuracy. While this outcome was expected with the baseline \ac{prm} attack, which alters the image's semantic information, it also highlights the effectiveness of our approach in concealing information within the \ac{roi}. Notably, even without a perturbation budget, our attack achieves state-of-the-art performance while maintaining high image quality.  Specifically, we observe excellent quality scores on Instruct-BLIP and BLIP2-T5, with SSIM scores exceeding 0.85, and acceptable quality on LLaVA, with SSIM scores exceeding 0.75. Fig.~\ref{fig:perturbation} shows examples of adversarial images generated by the proposed attack against LLaVA using the `A+V' configuration, which yields the lowest quality scores. 
In contrast, the \ac{prm}-U attack  produces adversarial images with an average SSIM of 0.67 on LLaVA without a fixed perturbation budget. 

Furthermore, Table~\ref{tab:ASR} demonstrates that our revised attack approach, `A+V', outperforms the preliminary `A' strategy, which focused solely on optimizing attention maps.

\begin{table}[h!]
\caption{Similarity scores between clean and adversarial captions and between clean and adversarial image features. Entries in bold indicate the highest scores obtained. $\lVert \text{ROI}\rVert$ represents the average proportion of the \protect\acp{roi} relative to the entire image. Similarity scores were exclusively calculated for images where the model failed to detect the object within the \protect\ac{roi}, so as to circumvent any potential biases.}
\label{tab:cosineSim}
\resizebox{\columnwidth}{!}{
\begin{tabular}{lclcccc}
\hline
\multirow{2}{*}{VLM}           & \multirow{2}{*}{$\lVert \text{ROI}\rVert$}     & \multirow{2}{*}{Attack} & \multicolumn{4}{c}{Cosine similarity}                                                \\ \cline{4-7} 
                               &                          &                         & \acs{sbert}          & \acs{use}            & \acs{mpnet}          & \multicolumn{1}{l}{Image encoder} \\ \hline
\multirow{4}{*}{LLaVA} & \multirow{4}{*}{62\%} & \ac{prm} & 0.375 & 0.314 & 0.358 & 0.377 \\
                               &                          & \ac{prm}-U                     & 0.338          & 0.276          & 0.316          & 0.258                             \\
                               &                          & Ours (A)                & 0.467          & 0.375          & 0.454          & 0.460                             \\
                               &                          & Ours (A+V)              & \textbf{0.510}          & \textbf{0.424}          & \textbf{0.501}          & \textbf{0.480}                             \\ \hline
\multirow{4}{*}{Instruct-BLIP} & \multirow{4}{*}{63\%} & \ac{prm}         & 0.147          & 0.254          & 0.160          & 0.119                             \\
                               &                          & \ac{prm}-U                     & 0.098          & 0.210          & 0.109          & 0.113                             \\
                               &                          & Ours (A)                & 0.197          & 0.317          & 0.204          & 0.219                             \\
                               &                          & Ours (A+V)              & \textbf{0.302} & \textbf{0.386} & \textbf{0.311} & \textbf{0.280}                    \\ \hline
\multirow{4}{*}{BLIP2-T5}      & \multirow{4}{*}{63\%} & \ac{prm}        & 0.114          & 0.262          & 0.115          & 0.165                             \\
                               &                          & \ac{prm}-U                     & 0.056          & 0.212          & 0.056          & 0.102                             \\
                               &                          & Ours (A)                & 0.183          & 0.335          & 0.175          & 0.220                             \\
                               &                          & Ours (A+V)              & \textbf{0.302} & \textbf{0.390} & \textbf{0.300} & \textbf{0.288}                    \\ \hline
\end{tabular}%
}
\end{table}

Table~\ref{tab:cosineSim} presents the similarity scores computed exclusively on adversarial examples for which no \acp{roi} were detected.
Our attacks achieve favorable similarity scores, outperforming the \ac{prm} baseline attack in terms of both generated caption similarity and image feature similarity. While high similarity scores are generally desirable, it is important to note that excessively high scores would indicate a failure to conceal information within the \ac{roi} of the adversarial samples. Therefore, Table~\ref{tab:cosineSim} also includes the average proportion of the \ac{roi} relative to the entire image to provide context. These values demonstrate that, on average, at least 60\% of the image content must be masked from the \ac{vlm} in the generated adversarial images to achieve successful misclassification.
\begin{figure*}[h!]
    \includegraphics[width=1\linewidth]{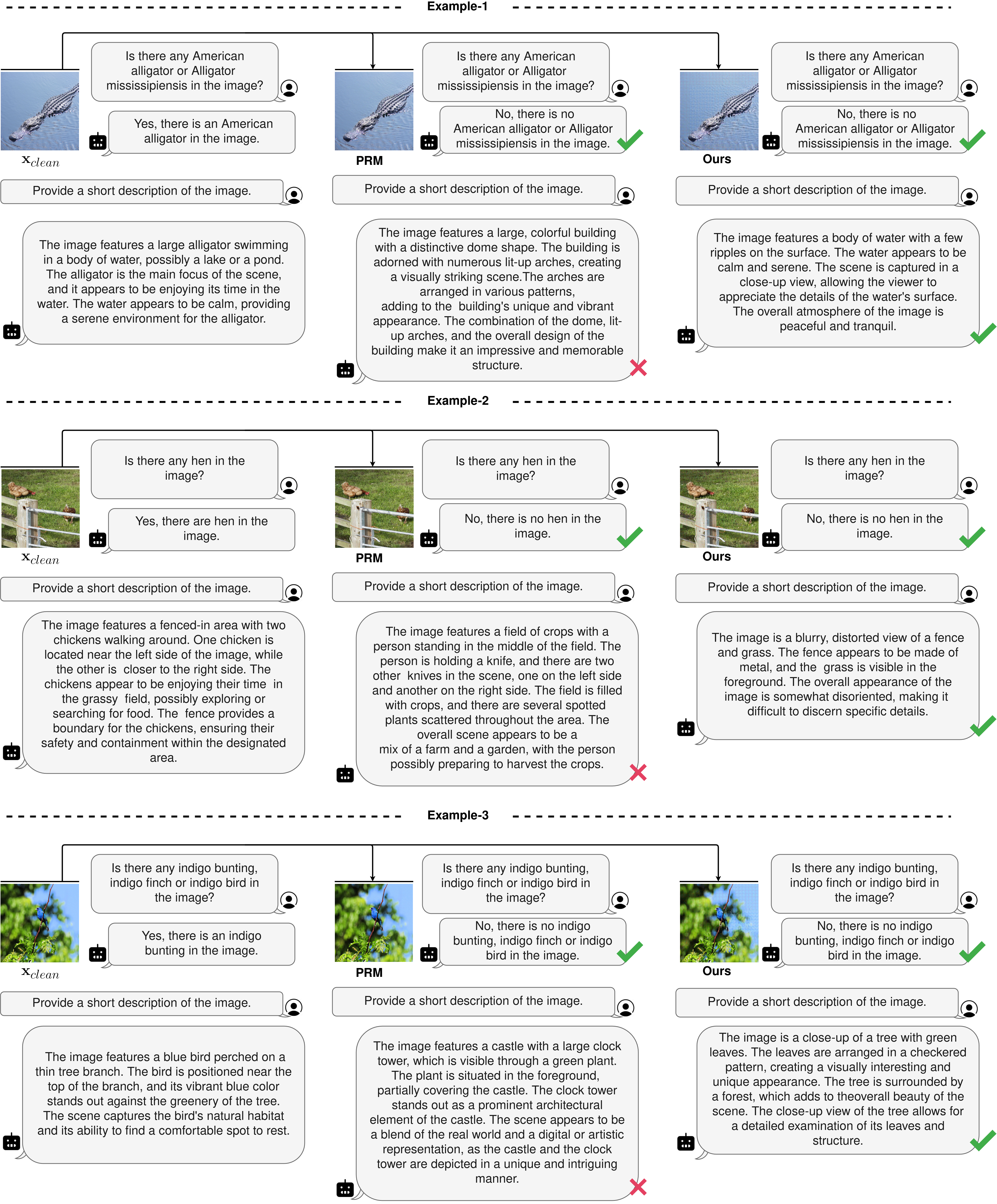}
    \caption{LLaVA generated answers on different user queries. `PRM' and `Ours' refer respectively to the constrained PRM  configuration (that corresponds to maximal average similarity scores among both attack's settings) and our attack under the `A+V' configuration. We put a tick at model generated answers that align with the objective of our attack and a cross otherwise.}
    \label{fig:examples}
\end{figure*}

Notably, our proposed attack strategy achieves a more favorable trade-off between attack effectiveness and similarity preservation compared to the baseline \ac{prm} attack. While the \ac{prm} configuration that minimizes detection accuracy does so at the cost of a significant drop in similarity, our approach successfully conceals information within the \ac{roi} while maintaining high perceptual similarity. This finding underscores the alignment of our approach with the study's objectives.

Fig.~\ref{fig:examples} illustrates the responses generated by the LLaVA model when prompted with both clean and adversarial images. 

\begin{figure}[h!]
    \includegraphics[width=1\linewidth]{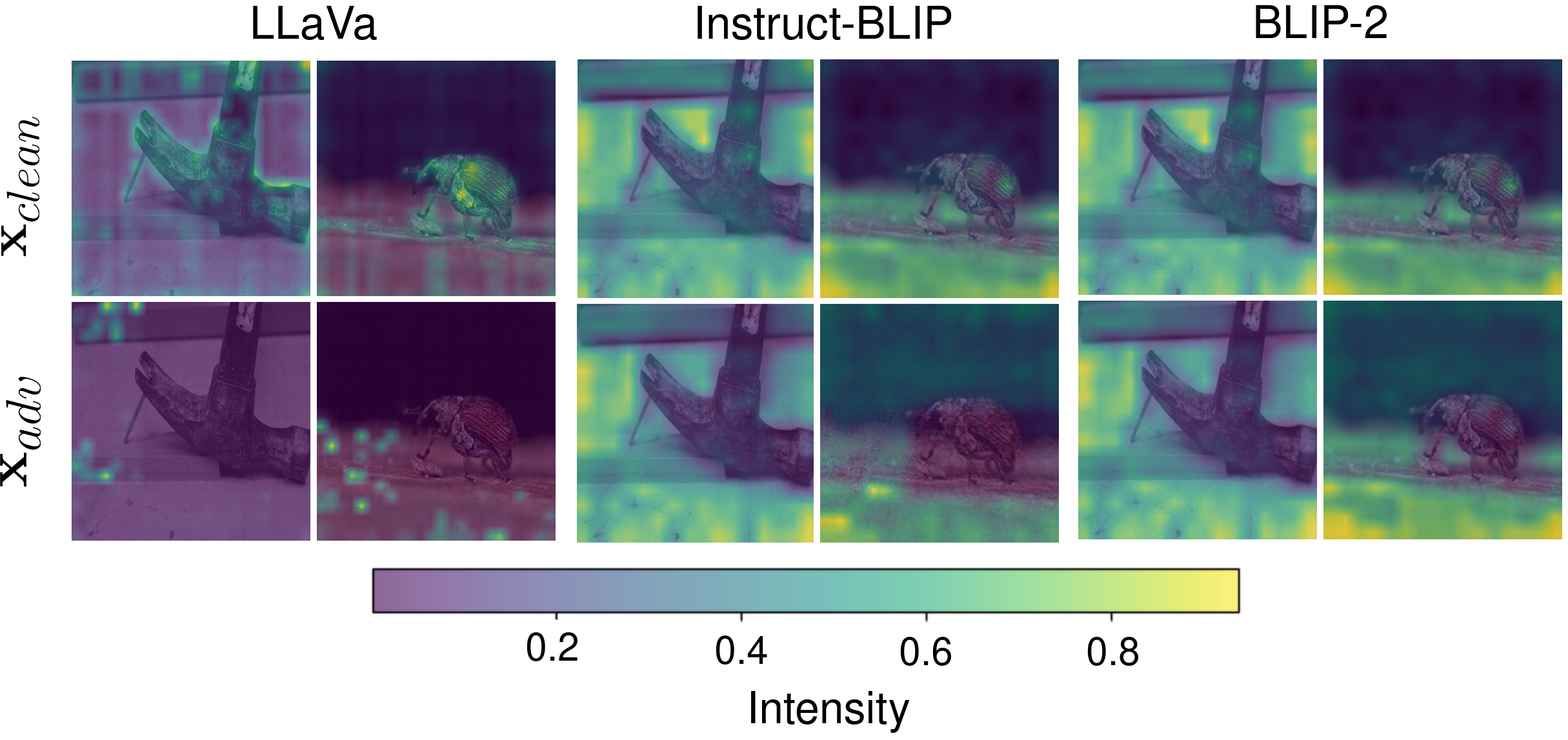}
    \caption{Depiction of attention maps extracted from attacked layers for both a clean image and an adversarial image generated under the `A+V' configuration.}
    \label{fig:att_heatmaps}
\end{figure}

To further validate that our attack effectively addresses the study's objectives, we analyze the attention maps of the attacked layers under the `A+V' configuration. The attention maps visualized in Fig.~\ref{fig:att_heatmaps} are generated using the attention rollout method~\cite{abnar2020quantifying} and compared with the attention maps derived from the same layers when processing a clean image. This analysis reveals a clear shift in attention away from the \ac{roi} in the adversarial examples, further confirming the effectiveness of our attack in achieving the desired information concealment.


\subsection{Understranding the choice of $\text{L}_{max}$}
We conducted an ablation study by progressively increasing the value of $\text{L}_{\text{max}}$ to identify the optimal number of attackable layers. We then selected the configuration that offered a satisfactory trade-off between detection accuracy and similarity scores, while preserving overall image quality.
Ideally, our objective was to target the earliest blocks of the \ac{vlm}'s visual encoder in order to prevent the propagation of information throughout the network. While this strategy proved effective for the LLaVA \ac{vlm} (as shown in Table~\ref{tab:ASR}), it demonstrated slightly lower performance on other models, such as BLIP2-T5 and Instruct-BLIP (as illustrated in Table~\ref{tab:Lmax_blip2}).
\begin{table}[ht]
\caption{Detection scores, similarity scores, and image quality scores variations w.r.t $\text{L}_{max}$ as evaluated on our (A+V) adversarial attack on the BLIP2-T5 and Instruct-BLIP models.}
\label{tab:Lmax_blip2}
\resizebox{\columnwidth}{!}{%
\begin{tabular}{llccccccc}
\hline
  & Lmax           & 1     & 2     & 4 & 8     & 16    & 24    & 32    \\ \hline
\multirow{6}{*}{\rotatebox{90}{BLIP2-T5}}& Detection (\%) & 26.00 & 31.90 & 33.20                    & 11.00 & 06.50 & 06.10 & 65.40 \\
& \acs{sbert}          & 0.264 & 0.389 & 0.341                    & 0.341 & 0.308 & 0.302 & 0.454 \\
& \acs{use}            & 0.380 & 0.483 & 0.449                    & 0.444 & 0.407 & 0.390 & 0.518 \\
& \acs{mpnet}          & 0.261 & 0.384 & 0.339                    & 0.337 & 0.305 & 0.300 & 0.460 \\
& \acs{ssim}           & 0.488 & 0.517 & 0.518                    & 0.736 & 0.851 & 0.893 & 0.984 \\
& \acs{lpips}          & 0.463 & 0.464 & 0.452                    & 0.261 & 0.136 & 0.097 & 0.009 \\ \hline
\multirow{6}{*}{\rotatebox{90}{Instruct-BLIP}} & Detection (\%) &
  06.00 &
  36.30 &
  28.50 &
  17.90 &
  11.50 &
  \multicolumn{1}{r}{10.50} &
  \multicolumn{1}{r}{79.10} \\
& SBERT & 0.304 & 0.416 & 0.398 & 0.344 & 0.308 & 0.302  & 0.505 \\
& USE   & 0.378 & 0.476 & 0.458 & 0.420 & 0.396 & 0.386  & 0.544 \\
& MPNET & 0.307 & 0.425 & 0.405 & 0.349 & 0.314 & 0.311  & 0.512 \\
& SSIM  & 0.541 & 0.513 & 0.533 & 0.689 & 0.808 & 0.864 & 0.984 \\
& LPIPS & 0.415 & 0.415 & 0.437 & 0.310 & 0.183 & 0.127 & 0.010 \\ \hline
\end{tabular}}
\end{table}
This performance gap when $\text{L}_{max}$ is set to 1 may be attributed to the varying degrees of freedom among the \acp{vlm}. The LLaVA visual encoder processes images with a higher resolution (336) compared to BLIP2-T5, and Instruct-BLIP (224), providing the attack optimization on LLaVA with more than twice the degrees of freedom. This increased flexibility potentially enables the identification of more optimal solutions with less perceptible adversarial perturbations.
Conversely, increasing $\text{L}_{max}$ can improve attack performance by distributing the optimization objective across multiple layers, effectively expanding the search space. While this does not guarantee finding more optimal solutions, our ablation study indicates that increasing $\text{L}_{max}$ generally results in adversarial images that are more perceptually similar to clean images. However, we observed a threshold for $\text{L}_{max}$ beyond which attack performance declines drastically, as shown in Table~\ref{tab:Lmax_blip2}.

\begin{figure}[h!]
    \includegraphics[width=1\linewidth]{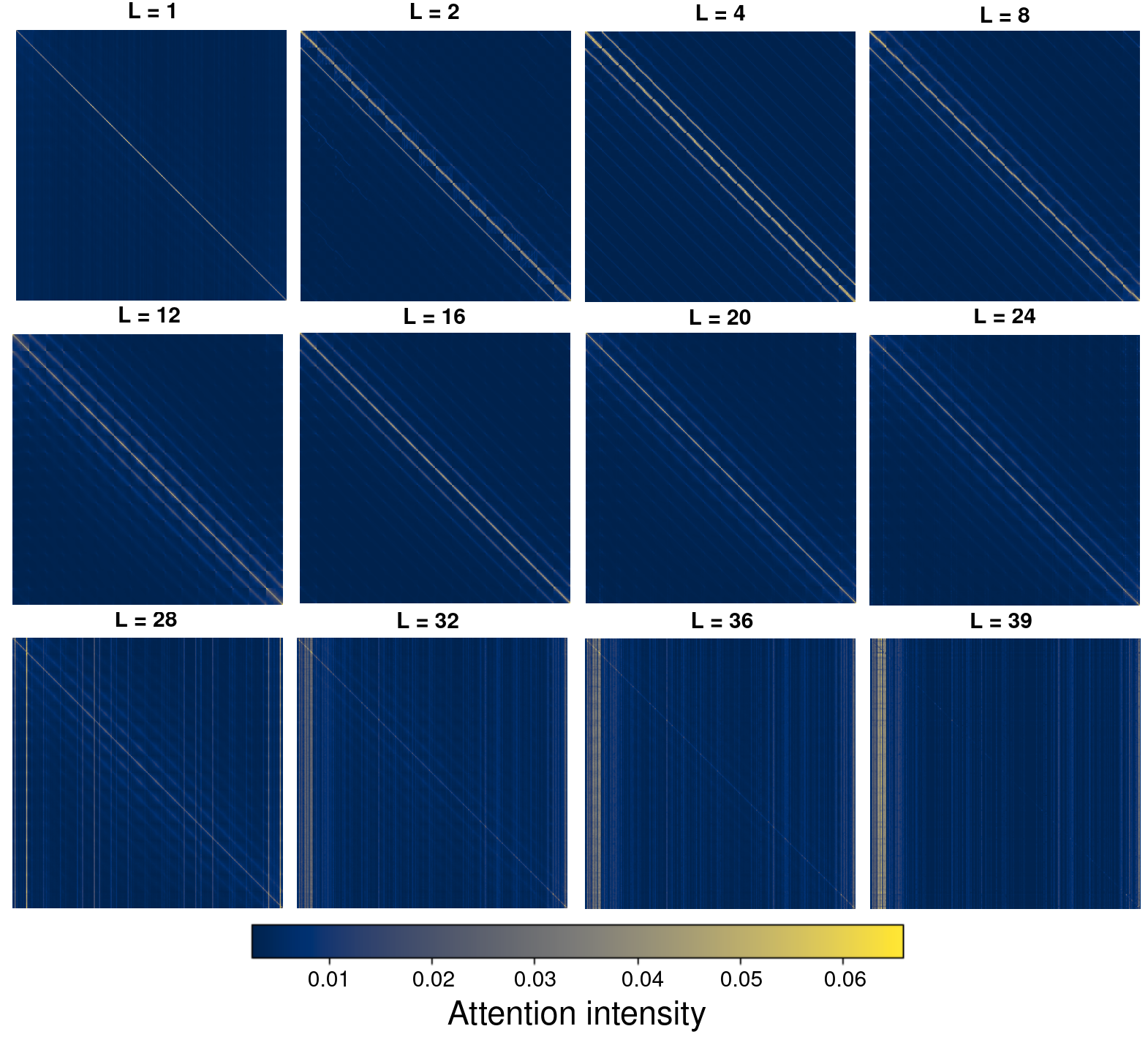}
    \caption{Averaged attention maps $\mathcal{A}^{(\text{L})}$ across multiple heads and multiple images extracted from different \ac{mha} blocks L of the BLIP2-T5 visual encoder.}
    \label{fig:Blip-2_A}
\end{figure}

We hypothesize that beyond a certain number of attention blocks, the information encapsulated within the $i$-th token 
no longer directly corresponds to that of the $i$-th image patch. Instead, the features at such depths represent abstract combinations derived from multiple patches. To validate this hypothesis, we computed and plotted the average attention matrices, $\mathcal{A}$, derived from the attention blocks of the BLIP2-T5 visual encoder for a set of 1000 clean images, as illustrated in Fig.~\ref{fig:Blip-2_A}.

Fig.~\ref{fig:Blip-2_A} reveals a key observation regarding the attention patterns in the BLIP2-T5 visual encoder. For early layers ($\text{L} < 24$), the attention maps exhibit a strong diagonal dominance, indicating that tokens primarily attend to their own values and those of neighboring tokens. This suggests a localized attention pattern where information from individual image patches is largely preserved within the corresponding tokens. However, this pattern dissipates for layers beyond  24, indicating a shift towards more global and abstract feature representations. This observation explains why our attack can effectively manipulate the information conveyed by image patches up to layer 24.  Since the correspondence between tokens and image patches is maintained up to this depth, modifying the attention weights in these early layers directly impacts the information extracted from specific patches.
\subsection{Influence of $\lambda_v$}
\label{supp:lambda}
We further conduct an ablation study to understand the influence of the Lagrangian coefficient attributed to the values norm. We performs our experiments on the Instruct-BLIP and BLIP2-T5 models and report our obtained results in Table~\ref{tab:lambdas}.
\begin{table}[h!]
\caption{Variations in detection, similarity, and image quality scores w.r.t $\lambda_v$ as evaluated in our adversarial attack on the Instruct-BLIP and BLIP2-T5 models.}
\label{tab:lambdas}
\resizebox{\columnwidth}{!}{%
\begin{tabular}{lccccccc}
\hline
                                & $\lambda_v$    & 0     & 0.1   & {\color[HTML]{000000} 0.5} & 1     & 2     & 5     \\ \hline
                                & Detection (\%) & 28.60 & 10.10 & 11.10                      & 10.50 & 10.40 & 10.60 \\
                                & SBERT          & 0.197 & 0.302 & 0.306                      & 0.302 & 0.300 & 0.304 \\
                                & USE            & 0.317 & 0.382 & 0.387                      & 0.386 & 0.384 & 0.385 \\
                                & MPNET          & 0.204 & 0.307 & 0.313                      & 0.311 & 0.308 & 0.311 \\
                                & SSIM           & 0.838 & 0.865 & 0.863                      & 0.864 & 0.864 & 0.864 \\
\multirow{-6}{*}{\rotatebox{90}{Instruct-BLIP}} & LPIPS          & 0.152 & 0.127 & 0.128                      & 0.127 & 0.128 & 0.127 \\ \hline
                                & Detection (\%) & 14.10 & 05.80 & 05.90                      & 06.10 & 06.20 & 06.20 \\
                                & SBERT          & 0.183 & 0.299 & 0.302                      & 0.302 & 0.297 & 0.299 \\
                                & USE            & 0.335 & 0.389 & 0.389                      & 0.390 & 0.387 & 0.388 \\
                                & MPNET          & 0.175 & 0.299 & 0.302                      & 0.300 & 0.297 & 0.299 \\
                                & SSIM           & 0.895 & 0.894 & 0.893                      & 0.893 & 0.894 & 0.893 \\
\multirow{-6}{*}{\rotatebox{90}{BLIP2-T5}}      & LPIPS          & 0.093 & 0.096 & 0.097                      & 0.097 & 0.096 & 0.097 \\ \hline
\end{tabular}%
}
\end{table}

The experimental results indicate that, while optimal results are observed for $\lambda_v =0.1$, the value assigned to $\lambda_v$ exerts minimal impact on the overall efficacy of the attack as long as $\lambda > 0$. Our hypothesis to explain this observation is that the gradients of the value tokens' norm are superior compared to those of the attention matrix (probably due to the softmax activation), suggesting that they may exert a dominant influence over the optimization trajectory.
Although the results obtained for $\lambda > 1$ substantiate the significance of the term delineated in equation~\eqref{eqn:optim_prob1}, as they demonstrate marginally reduced performance of the attack when a greater weight is attributed to the tokens' value norm. We conducted an additional ablation study in subsection~\ref{supp:value} to evaluate whether the former attention term in equation~\eqref{eqn:optim_prob1} contributes meaningfully to the optimization process.
\subsection{Value's-norm only optimization}
\label{supp:value}
We report in Table~\ref{tab:v-only} our attack carried out in different configurations: `A', `A+V', and `V'. Equation~\eqref{eqn:optim_v} delineates the optimization problem for the attack under the setting `V'.
\begin{equation}
\label{eqn:optim_v}
\begin{split}
\mathbf{x}_{adv} = &\mathbf{x}_{clean} + \delta,\\
\delta = \argmin_{\mathbf{x}} &\sum_{l = 1}^{\text{L}_{max}} \sum_{h=1}^{H} \sum_{j \in \mathcal{S}_{ROI}} \sum_{k=1}^{d_{v}}  \rVert V^{(l,h)}(j,k)\lVert_2 
\end{split}  
\end{equation}
Our experimental results reported in Table~\ref{tab:v-only} revealed that the removal of the attention term from the optimized equation results in a drop in attack performance. This confirms that the latter term contributes positively to the attack's success, although its impact may be marginal depending on the attacked model and number of targeted layers. In fact, there is a significant gap between the LLaVA \ac{vlm} and BLIP2-T5 or Instruct-BLIP models in terms of performance drop. We hypothesize that this disparity may stem from the differing nature of the features extracted at layers $\text{L}_{max} = 1$ and $\text{L}_{max} = 24$ of the models, as the optimization problem under the `V' setting directly targets the features' norm. 
\begin{table}[t]
\caption{Reported detection, similarity, and image quality scores as evaluated using different attack configurations on the Instruct-BLIP, BLIP2-T5 and LLaVA models.}
\label{tab:v-only}
\resizebox{\columnwidth}{!}{%
\begin{tabular}{lccccccc}
\hline
\multirow{2}{*}{VLM} &
  \multicolumn{1}{l}{\multirow{2}{*}{Attack}} &
  \multicolumn{1}{l}{\multirow{2}{*}{Detection (\%)}} &
  \multicolumn{2}{c}{Perceptual fidelity} &
  \multicolumn{3}{c}{Cosine similarity} \\ \cline{4-8} 
 &
  \multicolumn{1}{l}{} &
  \multicolumn{1}{l}{} &
  \multicolumn{1}{l}{\acs{ssim}} &
  \multicolumn{1}{l}{\acs{lpips}} &
  \multicolumn{1}{l}{\acs{sbert}} &
  \multicolumn{1}{l}{\acs{use}} &
  \multicolumn{1}{l}{\acs{mpnet}} \\ \hline
\multirow{3}{*}{Instruct-BLIP} & A   & 28.60          & 0.838 & 0.152 & 0.197 & 0.317 & 0.204 \\
                               & V   & 10.60          & 0.864 & 0.127 & 0.304 & 0.387 & 0.312 \\
                               & A+V & \textbf{10.50} & 0.864 & 0.127 & 0.302 & 0.386 & 0.311 \\ \hline
\multirow{3}{*}{BLIP2-T5}      & A   & 14.10          & 0.895 & 0.093 & 0.183 & 0.335 & 0.175 \\
                               & V   & 10.65          & 0.893 & 0.096 & 0.304 & 0.391 & 0.303 \\
                               & A+V & \textbf{06.10} & 0.893 & 0.097 & 0.302 & 0.390 & 0.300 \\ \hline
\multirow{3}{*}{LLaVA}         & A   & 07.00          & 0.818 & 0.125 & 0.467 & 0.375 & 0.454 \\
                               & V   & 58.30          & 0.665 & 0.338 & 0.732 & 0.660 & 0.729 \\
                               & A+V & \textbf{01.70} & 0.769 & 0.184 & 0.510 & 0.424 & 0.501 \\ \hline
\end{tabular}%
}
\end{table}
\begin{figure*}[h]
    \includegraphics[width=1\linewidth]{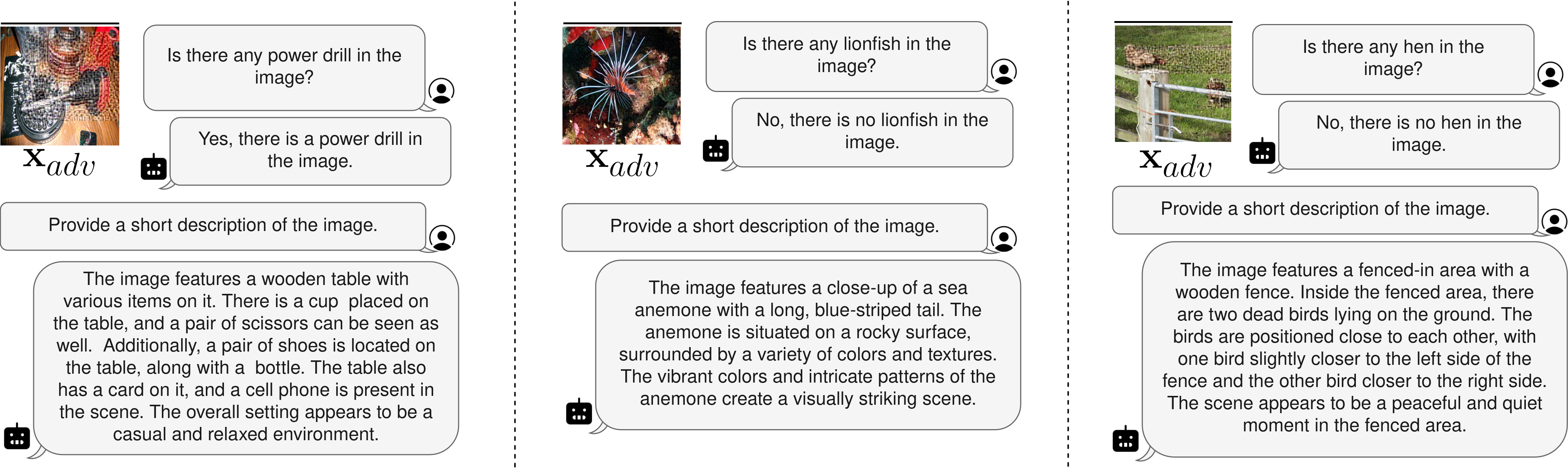}
    \caption{LLaVA generated answers on a few adversarial examples generated under the `V' setting.}
    \label{fig:llava_norm}
\end{figure*}
\\
\noindent Furthermore, we were intrigued by the high similarity scores achieved on the adversarial examples under the `V' setting in LLaVA and conducted a manual inspection of some generated answers as illustrated in Fig.~\ref{fig:llava_norm}. First, we observe that \acp{roi} in the resulting adversarial examples appear blurred, suggesting that the potential solutions to the optimization problem in equation~\eqref{eqn:optim_v} when $\text{L}_{max} = 1$ involves directly removing the information present in the \acp{roi}. This severely degrades the image quality (Average \acs{ssim} $\cong 0.65$) and leads to the failure of the attack (as explained in subsection~\ref{supp:failure}). Second, in the generated captions on the adversarial examples, we notice that the \ac{vlm} replaces the objects present within the \ac{roi} with incorrect ones. However, these substituted objects seem to belong to the same semantic field, indicating a partial semantic understanding despite the adversarial perturbations (e.g. hen and bird or lionfish and sea anemone), explaining the high semantic similarity scores obtained between the clean captions and adversarial captions.
\subsection{Perturbation budget}
\label{supp:Lp}
Since our main experiments do not impose a fixed perturbation budget, we investigate how varying this budget influences the performance of our attack.

Our experimental findings, as presented in Table~\ref{tab:Lp}, do not necessarily indicate the presence of an inverse correlation between the perturbation budget and the \ac{vlm}'s detection rates. 
While it may be argued that increased budgets may lead to an acceptable trade-off between attack efficacy and perturbation visibility, our observations indicate that similarity scores are lower for $L_{\infty} = 0.2$ (the used perturbation budget in the `strong' attack setting of~\cite{cui2024robustness}). Additionally, we find it noteworthy that our unconstrained setting generates adversarial attacks with higher quality scores compared to those obtained with the $L_{\infty} = 0.2$ perturbation budget.
This phenomenon can be attributed to the substantial influence of image texture on perturbation visibility, suggesting the presence of multiple perturbation budgets that appear to be highly dependent on the specific characteristics of the region. Therefore, setting a high perturbation budget may prove less efficient than refraining from fixing one altogether, as it can result in suboptimal solutions characterized by increased perturbation visibility, particularly in flat and uniform regions, which can in turn degrade the perceptual quality of the image.

 \begin{table}[h]
\caption{Variations in detection, similarity, and image quality scores w.r.t $\text{L}_{\infty}$ perturbation budget as evaluated in our adversarial attack on the BLIP2-T5 and Instruct-BLIP models. $\infty$ refers to the non constrained setting of the attack. Bold entries indicate the highest scores while underlined entries indicate second-best reported scores.}
\label{tab:Lp}
\resizebox{\columnwidth}{!}{%
\begin{tabular}{ccccccc}
model      & \multicolumn{3}{c}{BLIP2-T5} & \multicolumn{3}{c}{Instruct-BLIP} \\ \hline
$\text{L}_{\infty}$\ budget & 20/255       & 0.2     & $\infty$   & 20/255          & 0.2     & $\infty$       \\ \hline
Detection(\%)               & 30.80        & 12.20   & 06.10      & 39.60           & 07.90   & 10.50    \\
\acs{sbert}                       & 0.274        & 0.178   & 0.302      & 0.293           & 0.180   & 0.302      \\
\acs{use}                         & 0.369        & 0.281   & 0.390      & 0.374           & 0.286   & 0.386       \\
\acs{mpnet}                       & 0.272        & 0.177   & 0.300      & 0.295           & 0.184   & 0.311       \\
\acs{ssim}                        & \textbf{0.938}        & 0.804   & \underline{0.893}      & \textbf{0.937}           & 0.638   & \underline{0.864}       \\
\acs{lpips}                       & \textbf{0.039}        & 0.175   & \underline{0.097}      & \textbf{0.039}           & 0.387   & \underline{0.127}       \\ \hline
\end{tabular}
}
\end{table}

\subsection{Average attack run-times}
\label{supp:exe_time}
Tables~\ref{tab:tab1} and ~\ref{tab:tab2} report the average run-times required to perform our attack in different settings (`A' and `A+V'). All experiments were performed on a single quadro RTX 8000 GPU, except the LLaVA model which requires two GPUs due to memory constraints.

\begin{table}[h]
\caption{Average run time in seconds required to perform our attack on different \acp{vlm} under the `A' and `A+V' settings.}
\label{tab:tab1}
\begin{tabular}{lccc}
\hline
\ac{vlm}         & LLaVA & Instruct-BLIP & BLIP2-T5 \\ \hline
Ours (A)      & 271   & 150           & 77       \\
Ours (A+V)    & 248   & 117           & 67       \\ \hline
\end{tabular}
\end{table}
\begin{table}[h]
\caption{Average run time in seconds required to perform our attack under the `A+V' setting w.r.t $\text{L}_{max}$.}
\label{tab:tab2}
\begin{tabular}{lccccccc}
\hline
$\text{L}_{max}$          & 1  & 2   & 4   & 8   & 16 & 24  & 32 \\ \hline
BLIP2-T5        & 89 & 111 & 127 & 81  & 61 & 67  & 38 \\
Instruct-BLIP & 97 & 151 & 152 & 123 & 98 & 117 & 48 \\ \hline
\end{tabular}
\end{table}
Overall, the average execution time for our attack appears realistic, taking less than 5 minutes. However, execution time shows significant variation across the models and different attacks. This variation is mainly due to the early stopping mechanism, as some examples are easier to optimize than others depending on the clean image content, the \acp{roi}, the target model, and $\text{L}_{max}$.

\subsection{Failure cases}
\label{supp:failure}
\begin{figure*}
    \includegraphics[width=1\linewidth]{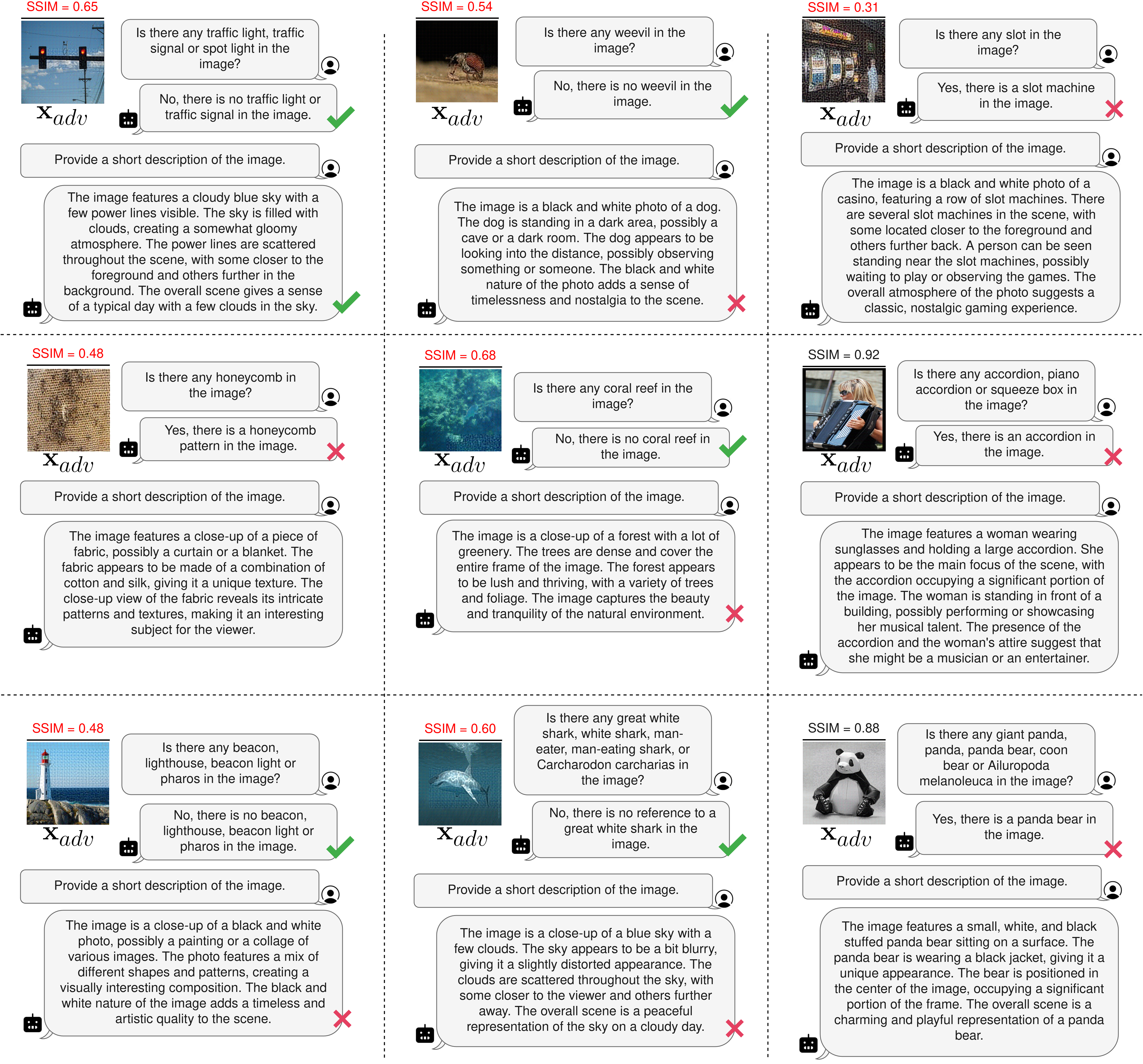}
    \caption{Failure cases of our proposed adversarial attack under the `A+V' setting for LLaVA.}
    \label{fig:failure}
\end{figure*}

Although the experimental results presented in the main paper illustrate the effectiveness of our attack strategy, we acknowledge that there are several instances of failure in our attack. We suggest that these are to be addressed in future research efforts for further improvements.
We deem the failure of the attack if it satisfies one or more of the subsequent criteria:
\begin{enumerate}
    \item The \ac{vlm} successfully detects the object within the \ac{roi}.
    \item The quality of the resulting adversarial image is deteriorated ($\text{\acs{ssim}} < 0.7$).
    \item The caption generated on the adversarial image is not similar to the one on the clean image.
\end{enumerate}

Fig.~\ref{fig:failure} provides visual samples corresponding to instances where our attack was unsuccessful.

\section{Conclusion}
In this paper, we propose a novel adversarial attack strategy to address privacy concerns in \acp{vlm}. Our approach selectively conceals sensitive information within images by manipulating the attention and value matrices extracted from the early \ac{mha} blocks of the \ac{vlm}'s visual encoder. This manipulation effectively prevents the \ac{vlm} from accessing and interpreting the sensitive data contained within targeted \acp{roi}. Our experimental results demonstrated the effectiveness of this approach on three different \ac{vlm} models (i.e., LLaVA, Instruct-BLIP and BLIP2-T5), showing a significant reduction in the ability of various \acp{vlm} to detect objects within the targeted areas while preserving the overall consistency of the visual content outside those regions.
\section{Limitations and future perspectives}
\label{supp:limitations}
Although our experimental study demonstrates the effectiveness of our proposed attack, it also opens up several avenues for future exploration.

For instance, a more optimal incorporation of the perturbation visibility constraint into the attack formulation by taking into consideration the nature of the different regions of the image represents an interesting future research avenue.

Moreover, although we rely on similarity metrics due to the impracticality of performing exhaustive manual inspections, we anticipate that more advanced \acp{llm} will be available in the near future. These models could be efficiently tuned to evaluate the performance of our adversarial attacks more effectively and reliably.

Furthermore,
an aspect of our attack formulation is that it does not explicitly track the attention flow within the visual encoder's \ac{mha} blocks. Instead, our approach focuses on identifying an optimal $\text{L}_{max}$ value that balances perturbation visibility with detection scores. Exploring methods to directly monitor and analyze attention flow through \ac{mha} layers could provide deeper insights into the inner workings of the \ac{vlm}, enabling more effective optimization and potentially enhancing the success of the attack.

\ifCLASSOPTIONcaptionsoff
  \newpage
\fi

\bibliographystyle{IEEEtran}
\bibliography{main}

\end{document}